\documentclass[twocolumn,aps,showpacs,preprintnumbers,superscriptaddress]{revtex4}
\usepackage{graphicx}
\usepackage{dcolumn}
\usepackage{bm}
\begin{document}

\title{Direct observation of higher-order whispering-gallery modes in VCSELs  at room temperature by embedding a defect-free surface micro-structure}

\author{Tsin-Dong Lee}
\affiliation{Graduate School of Optoelectronics, National Yunlin University of Science and Technology, Yunlin, 640 Taiwan}
\author{Chih-Yao Chen}
\affiliation{Institute of Photonics Technologies, National Tsing-Hua University, Hsinchu, 300 Taiwan}
\author{YuanYao Lin}
\affiliation{Institute of Photonics Technologies, National Tsing-Hua University, Hsinchu, 300 Taiwan}
\author{Chia-Yu Chang}
\affiliation{Graduate School of Optoelectronics, National Yunlin University of Science and Technology, Yunlin, 640 Taiwan}
\author{Ming-Chiu Chou}
\affiliation{Graduate School of Optoelectronics, National Yunlin University of Science and Technology, Yunlin, 640 Taiwan}
\author{Hung-Pin D. Yang}
\affiliation{Nanophotonic Center, Industrial Technology Research Institute, Hsinchu, 310 Taiwan}
\author{Te-ho Wu}
\affiliation{Institute of Material Sciences, National Yunlin University of Science and Technology, Yunlin, 640 Taiwan}
\author{Ray-Kuang Lee}
\affiliation{Institute of Photonics Technologies, National Tsing-Hua University, Hsinchu, 300 Taiwan}

\begin{abstract}
We propose and demonstrate a direct method to observe higher-order whispering-gallery modes in vertical cavity surface emitting lasers (VCSELs) at room temperature.
Instead of introducing any defect mode, we show that suppression of lower-order cavity modes can be achieved by destroying vertical reflectors with a surface micro-structure.
Up to the $23$rd azimuthal order whispering-gallery mode confined laterally by the native oxide layers is observed in experiments through collecting near-field radiation patterns.
Various vertical emission transverse modes are identified by the spectrum in experiments as well as numerical simulations.
\end{abstract} 

\maketitle
\date{\today}
Whispering-gallery modes (WGMs) are almost grazing incidence patterns confined by the total internal reflection at the interface \cite{WGM}.
With advantages of small mode volume and strong confinement WGMs have attracted much attention in photonics, quantum electrodynamics, and telecommunications, due to their potential application to enhance spontaneous emission and make threshold-less lasing. 
Fabrication of high-quality factor microdisk and microring has experimentally demonstrated WGMs with embedded emitters based on compound semiconductor lasers \cite{udisk, TDLee98}, and more recently on silicon-on-insulator materials \cite{usilicon}.
In the beginning, a waveguide is used to provide the confinement of light in the vertical direction for WGMs.
With optical output vertically emitted from the surface, vertical cavity surface emitting lasers (VCSELs) is a natural choice for lasers with transverse behavior of a WGM and a vertical emission \cite{Soda}.
Moreover, large area VCSELs also act as an interesting platform for studying optical pattern formation in mesoscopic system, such as scar modes \cite{KFHuang, Gensty} and cavity solitons \cite{Barland}. 

In recent years, with the advance of new fabrication technologies, it becomes more and more feasible to actually utilize one- or higher-dimensional periodic dielectric structures (or especially the photonic bandgap crystals) to modify the resonance modes in semiconductor lasers.
Combined with the bandgap effect and microcavities, ultra-high quality factor for a single defect semiconductor laser formed by a two-dimensional (2D) photonic crystal was reported \cite{Painter}.
Typically, WGMs surrounded in such various type of defect cavities are lowest-order modes due to that the defect geometry is with the same order of magnitude to the lasing wavelength.
Larger size microdisk  or microring have been introduced to excite higher-order WGMs \cite{Baba}.

\begin{figure}
\includegraphics[width=8.4cm]{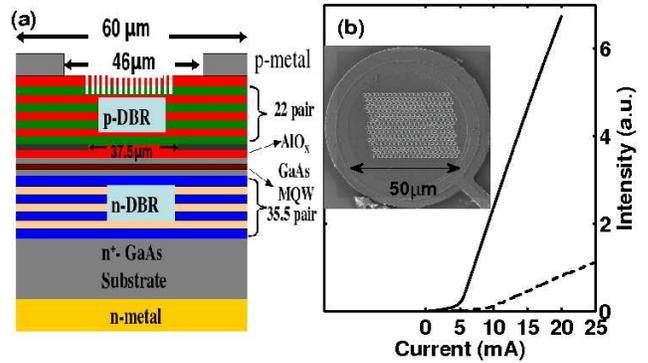}
\caption{Schematic diagram (a) and SEM image, inset in (b), of the device .
L-I curves (b) for our VCSEL without (solid-line) and with (dashed-line) a surface micro-structure.}
\label{Fig:F1}
\end{figure}

In this work, a 2D photonic crystal micro-structure is fabricated on a VCSEL surface \cite{Yokouchi} to investigate the transverse optical pattern formation by directly collecting near-field radiation intensity.
Instead of forming a defect cavity, we propose to use the surface structure as a deterioration mechanism for the desired lasing characteristics.
As the whole vertical emission window of VCSEL is destroyed by the surface micro-structure, we report the observation of higher-order WGMs confined laterally by the native oxide layer in a GaAs-based VCSEL at room temperature.
Through the suppression of lower-order cavity modes, up to the $23$rd azimuthal order whispering-gallery mode is observed both in experiments and simulations.
The differences of threshold current and emission spectra between the same VCSEL with and without a surface micro-structure are compared.
Moreover, we show that by increasing the injection current, different vertical emission transverse patterns, corresponding to the superposition of modes at different wavelengths, are identified both by the spectrum in experiments and by a 2D mode solver in simulations.
The experimental and numerical investigations in this work provide a new direction for studying the WGMs in semiconductor micro-structures.

The schematic diagram and SEM image of the micro-structured VCSEL used in our experiments are shown in Fig.\ref{Fig:F1}(a) and the inset in (b), respectively.
The epitaxial layers of the VCSELs are grown by MOCVD on a $n^+$-GaAs substrate, with GRINSCH active region formed by undoped triple-GaAs-AlGaAs quantum wells placed in one lambda cavity.
The upper and bottom DBRs in the vertical cavity consist of $22$ and $35.5$ pairs of Al$_{0.1}$Ga$_{0.9}$As/Al$_{0.9}$Ga$_{0.1}$As layers, respectively.
We introduce an oxide aperture to reduce the lateral optical loss and leakage current.
Then RIE is performed to define mesas with diameters of $68$ $\mu$m, where the Al$_{0.98}$Ga$_{0.02}$As layer within the Al$_{0.9}$Ga$_{0.1}$As confinement layers is selectively oxidized to AlO$_x$.
The oxidation depth is about $15$$\mu$m towards the center from the mesa edge so that the resulting oxide aperture is around $37.5$$\mu$m  in diameter. 
The $p$-contact ring with a inner diameter of $46$$\mu$m and a width of $7$$\mu$m is formed on the top of the $p$-contact layer.
The $n$-contact is formed at the bottom of the $n$-GaAs substrate.
Detail device parameter and lasing characteristics can be found in our previous publication on a similar VCSEL device but with different compound material \cite{TDLee07}.

As a comparison, the lasing characteristics of this VCSEL is measured before introducing any surface micro-structures.
Later on, the micro-structured surface pattern is defined by the focused ion beam.
A hexagonal lattice pattern without any defect is fabricated within the $p$-contact ring to introduce surface photonic crystal structures.
The lattice constant and diameter for this hexagonal photonic-crystal-structured VCSEL is $2$ and $1$$\mu$m, respectively.
Moreover, the depth for the surrounding holes is  only $0.2$$\mu$m in order to destroy the vertical reflector only, but not to introduce any photonic bandgap effect.
\begin{figure} 
\includegraphics[width=8.4cm]{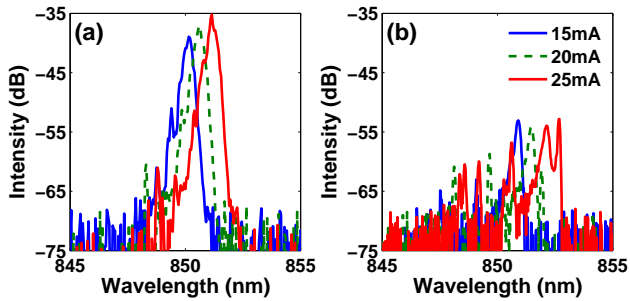}
\caption{The spectra of the VCSEL without (a) and with (b) a surface micro-structure at different injection currents.}
\label{Fig:F2}
\end{figure}
\begin{figure} 
\includegraphics[width=8.0cm]{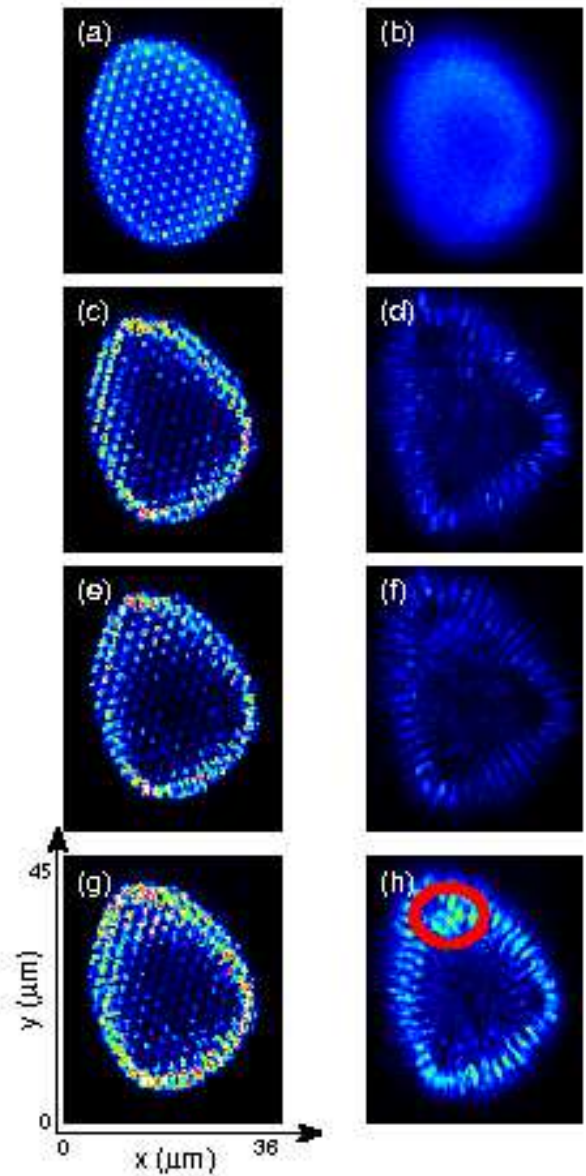}
\caption{Near field intensity distributions on the surface of aperture (left column) and emission window (right column) of our micro-structured VCSEL at different injection currents, i.e. $9$mA (a, b), $15$mA (c, d), $20$mA (e, f), and $25$mA (g, h).}
\label{Fig:F3}
\end{figure}

Fig.\ref{Fig:F1}(b) shows the L-I curve, light versus current, of the VCSEL without and with a photonic crystal structure on it.
The threshold currents for lasing operation are about $5$mA and $10$mA before and after introducing the surface micro-structure.
The increment of the threshold current is expected since that there is no defect mode induced.
After turning on, the emission spectra of our VCSEL without and with surface micro-structures at different operation currents are shown in Fig.\ref{Fig:F2} for a comparison.
In contrast to unstructured VCSELs, significant side modes appear in the shorter wavelengths.
Like a usual VCSEL without any surface structures, the main lasing peak in the spectrum for our micro-structured VCSEL also has the same tendency to shift to longer wavelength as the change of the refractive index induced by the injection currents.

Next, we measure the near-field electromagnetic intensity distribution at a fixed injection current.
The measured field intensity pattern is performed at different distances along the vertical direction, i.e. on the surface of aperture and emission window.
While the VCSEL is operated below threshold, for example, at the current of $9$mA, it can be seen clearly in Fig.\ref{Fig:F3}(a) that spontaneous emission pattern just reflects the surface structure and the lateral cavity defined by the oxide layer in the VCSEL.
The VCSEL is operated below the threshold condition now.
And at the emission window, as shown in Fig.\ref{Fig:F3}(b), we can see nothing but a uniform distribution of spontaneous emission pattern as a large area VCSEL \cite{Deng}.
In this case the designed surface photonic crystal structure has no effect on the lasing characteristics for only shallow holes are fabricated.
\begin{figure} 
\includegraphics[width=6.0cm]{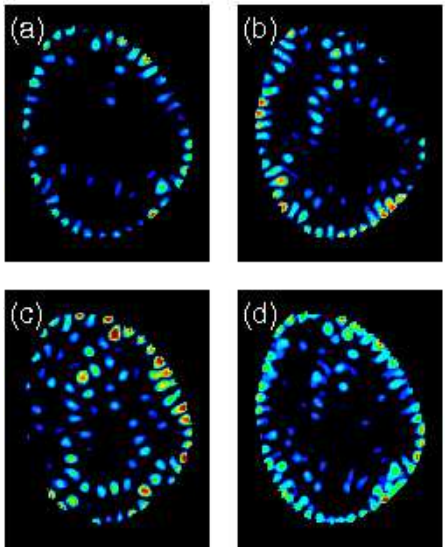}                        
\caption{Simulations of the field distribution of a surface structured VCSEL with the lateral boundary defined by an oxide layer. The intensity distributions for the eigen-modes at wavelengths of $852.811$nm (a), $853.062$nm (b), and $853.208$nm (c). The superposition of these three eigen-modes is shown in (d).}
\label{Fig:F4}
\end{figure}

As the injection current increases, the VCSEL is operated above threshold and begins to lase.
Owing to the destruction of top DBR reflector induced by the surface micro-structure, lower-order cavity modes are suppressed and higher-order WGMs have a chance to lase.
In Fig.\ref{Fig:F3}(c-d), operated at $15$mA a clear high-order WGM with multiple lobes is directly observed at room temperature with number of the azimuthal order to $23$.
When operated at higher current, i.e. $20$mA, in additional to the surrounding WGM scar-like patterns appear in the center region, as shown in Fig.\ref{Fig:F3}(e-f).
By increasing the current to $25$mA, a strongly localized field pattern is shown in the marked region in Fig.\ref{Fig:F3}(h).
At this moment, the spectrum of emitting profile contains multiple lasing peaks, as shown in Fig.\ref{Fig:F2}(b).

To verify the experimental observation of higher-order WGMs in such a defect-free micro-structured VCSEL, we perform a simple 2D model with a finite-element method to calculate cavity eigen-modes.
The lateral geometry defined by the native oxide layer is drawn according to the observed spontaneous emission pattern below the threshold current.
Then photonic crystal structures are embedded with the real lattice geometries.
The effective refractive indices are assumed to be $1$ in holes and $3.49$ in the surrounding $SiO_2$ oxide layer.
The calculated eigen-mode for the wavelengths at $852.811$nm is shown in Fig.\ref{Fig:F4}(a), which is a WGM with the same number of lobes in the azimuthal direction as the experimental data.
Fig.\ref{Fig:F4}(b) and (c) show the eigen-modes at wavelengths of $853.062$nm and $853.208$nm, with scar-like patterns in the center region. 
The superposition of these three transverse modes, as shown in Fig.\ref{Fig:F4}(d), results a good agreement with the experimental observation in Fig.\ref{Fig:F3}(h), and explains why there are multiple lasing peaks in the spectrum.

In conclusion, with micro-structure patterns and near-field technologies we investigate the formation of transverse optical patterns in GaAs-based VCSELs.
Without introducing any defect mode, the shallow surface structure is used to ruin the vertical reflector for the lower-order cavity modes.
At the expense of higher threshold current, a native oxide laterally confined whispering-gallery mode in VCSELs can be directly observed at room temperature. 
Up to the $23$rd azimuthal order vertical emission WGM is reported by collecting the near-field intensity on the surfaces of aperture and emission window.
The observed superposition of multiple lasing modes are identified by three transverse pattens with a direct numerical simulation. 
The experimental observations and the simulation results provide an alternative but easy approach to access WGMs in VCSELs at room temperature.

This work is partly supported by the National Science Council of Taiwan with contrasts NSC 95-2112-M-224-001, NSC 96-2112-M-224-001, NSC 95-2112-M-007-058-MY3 and NSC 95-2120-M-001-006.

\end{document}